# Acceleration and Strain-Rate Elastic Wave Equation and its Finite-Difference Solution for Distributed Acoustic Sensing Seismic Simulations

Saeed Izadian[1*]

[1]Institute of GeoEnergy Engineering, Heriot-Watt University, Edinburgh, UK, EH144AS

E-Mail: s.izadian@hw.ac.uk

## Abstract

Distributed Acoustic Sensing (DAS) is increasingly being adopted across various seismic disciplines due to its cost-effective acquisition capabilities and its high-resolution spatial and temporal sampling. Fiber-optic cables offer exceptional resistance in high pressure and temperature environments, facilitating both rapid-deployment and long-term continuous monitoring. Unlike traditional sensors, fibre optic cables record seismic vibrations as strain or strain-rate along the cables' longitudinal axis. To effectively integrate this data format, the governing equations for seismic wave propagation must be adjusted. This study introduces a formulation based on acceleration and strain-rate, analogous to the conventional velocity-stress framework used for describing elastic waves. The resulting system of first order, coupled partial differential equations closely mirrors existing velocity and stress models, allowing current theoretical frameworks and computational tools to be updated with minimal modification. Numerical validation using finite-difference simulations confirms that the proposed formulation produces strain-rate wavefields in complex media with accuracy equivalent to that of traditional velocity-stress methods, thereby effectively eliminating the requirement for data conversion or extra computations.

## 1. Introduction

Distributed Acoustic Sensing (DAS) is currently being tested and deployed across a diverse spectrum of seismic applications. Compared to conventional seismic sensors, fibre optic cables offer superior cost-efficiency and exceptional resistance in high pressure and temperature environments. Furthermore, they facilitate seismic data acquisition with significantly enhanced spatial and temporal sampling rate (Li et al. 2022). Consequently, the adoption of fibre optic technology has expanded rapidly in recent years in a wide range of disciplines such as vertical seismic profiling (Mateeva et al. 2017; Yurikov et al. 2022), micro-seismicity monitoring (Walter et al. 2020; Hudson et al. 2021), ambient noise interferometry (Nayak et al. 2021), time-lapse/4D seismic surveys (Harris et al. 2016; Kitawaki et al. 2023), and near-surface site characterization (Dou et al. 2017). Despite its advantages, DAS recorded data is fundamentally distinct from that acquired via conventional seismic sensors. These differences arise from two primary factors; first, the interrogation system records strain or strain-rate derived from phase changes in Rayleigh backscattering, rather than the discrete particle velocity or acceleration measured by geophones; second, due to the nature of the backscattering phenomenon, recordings are inherently restricted to the longitudinal axis (tangency) of the fibre-optic cable (Li et al. 2022). Consequently, established theoretical frameworks and practical processing workflows must be recalibrated to accurately account for these unique physical properties.

*Corresponding Author, Email: s.izadian@hw.ac.uk

A critical area requiring modification is the governing equation for elastic waves, which serves as the foundation for subsequent seismic imaging theories. Traditionally, the velocity-stress formulation for elastic media and the pressure wavefield formulation for acoustic media have been the standards, as they align directly with the measurements captured by conventional sensors. One method for adapting existing frameworks to DAS data involves deriving strain or strain-rate from particle displacement or velocity vectors (Eaid et al. 2018; Eaid et al. 2020; Wu, 2021; Lecoulant et al. 2023). However, this conversion introduces theoretical and computational overhead. Alternatively, one may convert recorded strain or strain-rate data into particle velocity, but this methodology is inherently susceptible to signal errors (Li et al. 2022; Zhao et al. 2025). To circumvent these complexities, a more efficient strategy is to derive governing equations that natively incorporate strain or strain-rate. Among the few studies on this approach is the stress and strain-rate formulation proposed by Zhao et al. (2025). Their approach results in a system of first- and second-order partial differential equations which complicates the implementation of numerical solvers and necessitates substantial modifications to established analytical tools and methodologies. Since the primary objective is the direct calculation of strain-rate tensor components, this study proposes an elastic wave equation formulated in terms of acceleration and strain-rate. The principal advantage of this framework is its mathematical similarity to the conventional velocity-stress formulation. Consequently, its implementation requires only minimal modifications to existing theoretical frameworks and computational workflows.

In the following, first, the theoretical framework of governing elastodynamics equations is reviewed. Subsequently, the acceleration and strain-rate formulation for elastic wave equation in anisotropic media is derived following a mathematical process analogous to the velocity-stress method. Following the elastic formulation, the acoustic counterpart of the acceleration and strain-rate formulation is derived and discussed. Afterwards, calculation of the tangential strain-rate component from the full tensor and fundamental principles of the finite-difference solution specifically tailored for this formulation, are detailed. Following the theory section, two numerical examples are presented to evaluate the accuracy of the proposed method relative to the velocity-stress formulation, highlighting its robust performance in complex media. Finally, the implications and conclusions of this study are provided.

## 2. Theory

The elastic wave equation is fundamentally derived from three constitutive elements. The primary component is Newton's law of motion, which establishes the relationship between applied forces, material density, and particle displacement. In the Cartesian coordinate system, the equations of motion are as follows (Gurtin 1982; Chapman 2004)

$$\rho\frac{\partial^2 u_x}{\partial t^2}=\frac{\partial\sigma_{xx}}{\partial x}+\frac{\partial\sigma_{xy}}{\partial y}+\frac{\partial\sigma_{xz}}{\partial z}+f_x,$$
$$\rho\frac{\partial^2 u_y}{\partial t^2}=\frac{\partial\sigma_{xy}}{\partial x}+\frac{\partial\sigma_{yy}}{\partial y}+\frac{\partial\sigma_{yz}}{\partial z}+f_y,$$
$$\rho\frac{\partial^2 u_z}{\partial t^2}=\frac{\partial\sigma_{xz}}{\partial x}+\frac{\partial\sigma_{yz}}{\partial y}+\frac{\partial\sigma_{zz}}{\partial z}+f_z, \tag{1}$$

where $\rho$ denotes the density, $u$s are the elements of the displacement vector, $\sigma$s are components of the stress tensor, and $f$s are the external forces triggering the motion. The second fundamental component of the elastic wave equation is the constitutive relationship, which establishes a linear mapping between the stress and strain tensors via the stiffness tensor. In this study, the stiffness tensor for a vertically transverse isotropic (VTI) medium serves as the foundational baseline. It should be noted that alternative formulations, such as isotropic or tilted transversely isotropic wave equations, can be systematically derived using the same mathematical framework. In its expanded form, the constitutive model for a VTI medium is expressed as following (Tsvankin 1996)

$$\sigma_{xx}=c_{11}\varepsilon_{xx}+c_{12}\varepsilon_{yy}+c_{13}\varepsilon_{zz},$$
$$\sigma_{yy}=c_{12}\varepsilon_{xx}+c_{11}\varepsilon_{yy}+c_{13}\varepsilon_{zz},$$
$$\sigma_{zz}=c_{13}\varepsilon_{xx}+c_{13}\varepsilon_{yy}+c_{33}\varepsilon_{zz},$$
$$\sigma_{xy}=2c_{66}\varepsilon_{xy},$$
$$\sigma_{xz}=2c_{44}\varepsilon_{xz},$$
$$\sigma_{yz}=2c_{44}\varepsilon_{yz}, \tag{2}$$

where $c$s are the elements of the stiffness tensor, and components of the strain tensor are denoted by $\varepsilon$s. To enhance readability, the stiffness tensor components are indexed using Voigt notation (Gurtin 1982; Tsvankin 1996). The explicit matrix representation of the stiffness tensor, along with definitions relative to the seismic properties, is detailed in the Appendix A. Given that seismic wave propagation induces only minor perturbations within subsurface rocks, the infinitesimal strain model provides a sufficient framework for the elastic wave equation, which is defined in terms of the displacement gradient as follows (Gurtin 1982; Chapman 2004)

$$\varepsilon_{xx}=\frac{\partial u_x}{\partial x},$$
$$\varepsilon_{yy}=\frac{\partial u_y}{\partial y},$$
$$\varepsilon_{zz}=\frac{\partial u_z}{\partial z},$$
$$\varepsilon_{xy}=\frac{1}{2}\left(\frac{\partial u_x}{\partial y}+\frac{\partial u_y}{\partial x}\right),$$
$$\varepsilon_{xz}=\frac{1}{2}\left(\frac{\partial u_x}{\partial z}+\frac{\partial u_z}{\partial x}\right),$$
$$\varepsilon_{yz}=\frac{1}{2}\left(\frac{\partial u_y}{\partial z}+\frac{\partial u_z}{\partial y}\right). \tag{3}$$

The three components detailed in this section are integrated to derive various formulations of the elastic wave equation. The following analysis, first, reviews the standard velocity-stress formulation, a widely adopted framework in seismic modeling. Subsequently, utilizing the same fundamental principles, acceleration and strain-rate formulation is derived. Finally, its acoustic counterpart alongside the corresponding numerical solutions is presented and discussed.

### 2.1. Elastic Wave Equation with Velocity and Stress Components

In order to derive the velocity-stress wave equation, first, the displacement elements in equation (1) are replace based upon the definition of the particle velocity as following

$$v_i = \frac{\partial u_i}{\partial t}, i = x, y, z. \tag{4}$$

Subsequently, the strain components defined in Eq. (3) are substituted into the constitutive relation in Eq. (2). By applying a partial time derivative to both sides of the resulting expression, a system of first order and coupled elastic wave equations are derived

$$\begin{aligned}
\rho \frac{\partial v_x}{\partial t} &= \frac{\partial \sigma_{xx}}{\partial x} + \frac{\partial \sigma_{xy}}{\partial y} + \frac{\partial \sigma_{xz}}{\partial z} + f_x, \\
\rho \frac{\partial v_y}{\partial t} &= \frac{\partial \sigma_{xy}}{\partial x} + \frac{\partial \sigma_{yy}}{\partial y} + \frac{\partial \sigma_{yz}}{\partial z} + f_y, \\
\rho \frac{\partial v_z}{\partial t} &= \frac{\partial \sigma_{xz}}{\partial x} + \frac{\partial \sigma_{yz}}{\partial y} + \frac{\partial \sigma_{zz}}{\partial z} + f_z, \\
\frac{\partial \sigma_{xx}}{\partial t} &= c_{11} \frac{\partial v_x}{\partial x} + c_{12} \frac{\partial v_y}{\partial y} + c_{13} \frac{\partial v_z}{\partial z}, \\
\frac{\partial \sigma_{yy}}{\partial t} &= c_{12} \frac{\partial v_x}{\partial x} + c_{11} \frac{\partial v_y}{\partial y} + c_{13} \frac{\partial v_z}{\partial z}, \\
\frac{\partial \sigma_{zz}}{\partial t} &= c_{13} \frac{\partial v_x}{\partial x} + c_{13} \frac{\partial v_y}{\partial y} + c_{33} \frac{\partial v_z}{\partial z}, \\
\frac{\partial \sigma_{xy}}{\partial t} &= c_{66} \left( \frac{\partial v_x}{\partial y} + \frac{\partial v_y}{\partial x} \right), \\
\frac{\partial \sigma_{xz}}{\partial t} &= c_{44} \left( \frac{\partial v_x}{\partial z} + \frac{\partial v_z}{\partial x} \right), \\
\frac{\partial \sigma_{yz}}{\partial t} &= c_{44} \left( \frac{\partial v_y}{\partial z} + \frac{\partial v_z}{\partial y} \right).
\end{aligned} \tag{5}$$

This formulation has gained traction across various seismic simulations and practical applications, primarily due to its compatibility with data recorded by standard seismic sensors, such as geophones. Virieux (1986) derives equations in Eq. (5) and establishes a robust framework for their numerical solution via the staggered-grid finite-difference technique. Followingly, numerous variations of this formulation and a diverse array of finite-difference schemes have been developed (e.g., Levander 1988; Saenger and Bohlen 2004; Moczo et al. 2007). In the subsequent section, a similar approach is employed to derive an analogous set of equations, formulated instead in terms of acceleration and strain-rate.

### 2.2. Elastic Wave Equation with Acceleration and Strain-Rate Components

To derive the wave equation in terms of acceleration and strain-rate, a temporal derivative is first applied to both sides of Eq. (1). The displacement components are subsequently replaced using the definition of particle acceleration in the following relationship

$$a_i = \frac{\partial v_i}{\partial t} = \frac{\partial^2 u_i}{\partial t^2}\ , i = x, y, z. \tag{6}$$

Hence, variations of the particle acceleration are related to the rate of stresses applied to that particle as following

$$\begin{aligned}
\rho \frac{\partial a_x}{\partial t} &= \frac{\partial \dot{\sigma}_{xx}}{\partial x} + \frac{\partial \dot{\sigma}_{xy}}{\partial y} + \frac{\partial \dot{\sigma}_{xz}}{\partial z} + \frac{\partial f_x}{\partial t}, \\
\rho \frac{\partial a_y}{\partial t} &= \frac{\partial \dot{\sigma}_{xy}}{\partial x} + \frac{\partial \dot{\sigma}_{yy}}{\partial y} + \frac{\partial \dot{\sigma}_{yz}}{\partial z} + \frac{\partial f_y}{\partial t}, \\
\rho \frac{\partial a_z}{\partial t} &= \frac{\partial \dot{\sigma}_{xz}}{\partial x} + \frac{\partial \dot{\sigma}_{yz}}{\partial y} + \frac{\partial \dot{\sigma}_{zz}}{\partial z} + \frac{\partial f_z}{\partial t}.
\end{aligned} \tag{7}$$

To distinguish the temporal rates of change for stress and strain, this study adopts the dot notation, denoted as $\dot{\sigma}$ and $\dot{\varepsilon}$, respectively. The derivation proceeds in three primary stages. First, a temporal derivative is applied to both sides of the constitutive relation in Eq. (2) to establish the link between stress and strain-rates. Second, the resulting stress rates are substituted into Eq. (7) to form the initial component of the acceleration and strain-rate system where it is critical to note that the temporal derivative is simultaneously applied to the source term as well. Finally, a second-order temporal derivative is applied to the strain-displacement relation in Eq. (3) to derive the second component of the system. The integration of these two sets of equations yields the following acceleration and strain-rate elastic wave equation

$$\begin{aligned}
\rho \frac{\partial a_x}{\partial t} &= \frac{\partial}{\partial x}\left(c_{11}\dot{\varepsilon}_{xx} + c_{12}\dot{\varepsilon}_{yy} + c_{13}\dot{\varepsilon}_{zz}\right) + 2\frac{\partial}{\partial y}\left(c_{66}\dot{\varepsilon}_{xy}\right) + 2\frac{\partial}{\partial z}\left(c_{44}\dot{\varepsilon}_{xz}\right) + \frac{\partial f_x}{\partial t}, \\
\rho \frac{\partial a_y}{\partial t} &= 2\frac{\partial}{\partial x}\left(c_{66}\dot{\varepsilon}_{xy}\right) + \frac{\partial}{\partial y}\left(c_{12}\dot{\varepsilon}_{xx} + c_{11}\dot{\varepsilon}_{yy} + c_{13}\dot{\varepsilon}_{zz}\right) + 2\frac{\partial}{\partial z}\left(c_{44}\dot{\varepsilon}_{yz}\right) + \frac{\partial f_y}{\partial t}, \\
\rho \frac{\partial a_z}{\partial t} &= 2\frac{\partial}{\partial x}\left(c_{44}\dot{\varepsilon}_{xz}\right) + 2\frac{\partial}{\partial y}\left(c_{44}\dot{\varepsilon}_{yz}\right) + \frac{\partial}{\partial z}\left(c_{13}\dot{\varepsilon}_{xx} + c_{13}\dot{\varepsilon}_{yy} + c_{33}\dot{\varepsilon}_{zz}\right) + \frac{\partial f_z}{\partial t}, \\
\frac{\partial \dot{\varepsilon}_{xx}}{\partial t} &= \frac{\partial a_x}{\partial x}, \\
\frac{\partial \dot{\varepsilon}_{yy}}{\partial t} &= \frac{\partial a_y}{\partial y}, \\
\frac{\partial \dot{\varepsilon}_{zz}}{\partial t} &= \frac{\partial a_z}{\partial z}, \\
\frac{\partial \dot{\varepsilon}_{xy}}{\partial t} &= \frac{1}{2}\left(\frac{\partial a_x}{\partial y} + \frac{\partial a_y}{\partial x}\right), \\
\frac{\partial \dot{\varepsilon}_{xz}}{\partial t} &= \frac{1}{2}\left(\frac{\partial a_x}{\partial z} + \frac{\partial a_z}{\partial x}\right), \\
\frac{\partial \dot{\varepsilon}_{yz}}{\partial t} &= \frac{1}{2}\left(\frac{\partial a_y}{\partial z} + \frac{\partial a_z}{\partial y}\right).
\end{aligned} \tag{8}$$

As established in the preceding section, the proposed formulation maintains a high degree of structural symmetry with the standard velocity-stress equations. Both frameworks comprise a system of nine coupled, first-order partial differential equations, exhibiting nearly identical mathematical architectures. This structural correspondence offers significant advantages for practical implementation. Existing imaging theories and numerical solvers optimized for the velocity-stress formulation can be seamlessly adapted to accommodate the acceleration and strain-rate framework with minimal algorithmic modification.

### 2.3. From Elastic to Acoustic Wave Equation

Despite its inherent limitations, the acoustic approximation is widely adopted across numerous seismic applications (Tarantola 1984; Yilmaz 2001). This assumption significantly reduces the computational overhead by isolating the compressional (P-wave) components of the wavefield, thereby facilitating the characterization of P-wave related properties. The acoustic media, such as air or fluids, are characterized by a negligible shear modulus, $\mu$ (Gurtin 1982; Chapman 2004). In other words, an acoustic medium lacks the capacity to resist shear stress. Consequently, the acoustic counterpart of the elastic wave equation is derived by setting either the shear modulus or the S-wave velocity, $v_s$, to zero ($\mu = v_s = 0$), yielding the following system of equations

$$
\begin{aligned}
&\rho \frac{\partial a_x}{\partial t} = \frac{\partial}{\partial x}\left(c_{11}^p \dot{\varepsilon}_{xx} + c_{11}^p \dot{\varepsilon}_{yy} + c_{13}^p \dot{\varepsilon}_{zz}\right) + \frac{\partial f_x}{\partial t}, \\
&\rho \frac{\partial a_y}{\partial t} = \frac{\partial}{\partial y}\left(c_{11}^p \dot{\varepsilon}_{xx} + c_{11}^p \dot{\varepsilon}_{yy} + c_{13}^p \dot{\varepsilon}_{zz}\right) + \frac{\partial f_y}{\partial t}, \\
&\rho \frac{\partial a_z}{\partial t} = \frac{\partial}{\partial z}\left(c_{13}^p \dot{\varepsilon}_{xx} + c_{13}^p \dot{\varepsilon}_{yy} + c_{33}^p \dot{\varepsilon}_{zz}\right) + \frac{\partial f_z}{\partial t}, \\
&\frac{\partial \dot{\varepsilon}_{xx}}{\partial t} = \frac{\partial a_x}{\partial x}, \\
&\frac{\partial \dot{\varepsilon}_{yy}}{\partial t} = \frac{\partial a_y}{\partial y}, \\
&\frac{\partial \dot{\varepsilon}_{zz}}{\partial t} = \frac{\partial a_z}{\partial z}, \\
&\frac{\partial \dot{\varepsilon}_{xy}}{\partial t} = \frac{1}{2}\left(\frac{\partial a_x}{\partial y} + \frac{\partial a_y}{\partial x}\right), \\
&\frac{\partial \dot{\varepsilon}_{xz}}{\partial t} = \frac{1}{2}\left(\frac{\partial a_x}{\partial z} + \frac{\partial a_z}{\partial x}\right), \\
&\frac{\partial \dot{\varepsilon}_{yz}}{\partial t} = \frac{1}{2}\left(\frac{\partial a_y}{\partial z} + \frac{\partial a_z}{\partial y}\right),
\end{aligned}
\tag{9}
$$

where $c^p s$ are components of the stiffness tensor in an acoustic medium (see Appendix A). However, the above formulation is physically inconsistent as the shear strain components remain non-zero. While the constitutive relationship dictates that nullifying the S-wave velocity effectively eliminates shear stresses, this logic does not apply to shear strains. Consequently, Eq. (9) describes a theoretical medium where shear stresses are absent, yet shear strains can still be generated and propagated, which is a condition that is physically contradictory for an acoustic medium. To illustrate this inconsistency, as an example consider the evaluation of the shear stress component, $\sigma_{xy}$, within an isotropic medium, which is defined as

$$\sigma_{xy} = 2\mu\varepsilon_{xy}. \tag{10}$$

where Setting $\mu = 0$ necessitates that $\sigma_{xy} = 0$, independent of the magnitude of the shear strain, $\varepsilon_{xy}$. This decoupling explains the persistence of shear strain terms in Eq. (9) as the shear strains are not constrained to zero. To further evaluate this relationship, the constitutive model can be re-formulated through the compliance tensor, where the strain component is expressed as a function of the corresponding stress component as following

$$\varepsilon_{xy} = \frac{1}{2\mu}\sigma_{xy}. \tag{11}$$

In this context, simultaneously nullifying the shear modulus and the shear stress results in a mathematically indeterminate zero divided by zero expression. Consequently, to derive a physically and mathematically consistent acoustic wave equation within the acceleration and strain-rate framework, the shear strain components must be explicitly omitted. Applying this constraint yields the following system of equations

$$\begin{aligned}
&\rho\frac{\partial a_x}{\partial t} = \frac{\partial}{\partial x}\left(c_{11}^{p}\dot{\varepsilon}_{xx} + c_{11}^{p}\dot{\varepsilon}_{yy} + c_{13}^{p}\dot{\varepsilon}_{zz}\right) + \frac{\partial f_x}{\partial t},\\
&\rho\frac{\partial a_y}{\partial t} = \frac{\partial}{\partial y}\left(c_{11}^{p}\dot{\varepsilon}_{xx} + c_{11}^{p}\dot{\varepsilon}_{yy} + c_{13}^{p}\dot{\varepsilon}_{zz}\right) + \frac{\partial f_y}{\partial t},\\
&\rho\frac{\partial a_z}{\partial t} = \frac{\partial}{\partial z}\left(c_{13}^{p}\dot{\varepsilon}_{xx} + c_{13}^{p}\dot{\varepsilon}_{yy} + c_{33}^{p}\dot{\varepsilon}_{zz}\right) + \frac{\partial f_z}{\partial t},\\
&\frac{\partial \dot{\varepsilon}_{xx}}{\partial t} = \frac{\partial a_x}{\partial x},\\
&\frac{\partial \dot{\varepsilon}_{yy}}{\partial t} = \frac{\partial a_y}{\partial y},\\
&\frac{\partial \dot{\varepsilon}_{zz}}{\partial t} = \frac{\partial a_z}{\partial z}.
\end{aligned} \tag{12}$$

It is important to note that, consistent with the elastic case, the acoustic counterpart of the acceleration and strain-rate formulation maintains a high degree of structural symmetry with the velocity-stress acoustic variant. Consequently, this formulation inherits the same computational and implementation advantages, ensuring seamless integration into existing numerical frameworks and imaging workflows.

### 2.4. Finite-Difference Solution

The acceleration and strain-rate formulation is derived to closely mirror the conventional velocity-stress framework. Consequently, the established finite-difference techniques and numerical algorithms employed for velocity-stress solutions are directly transferable to the acceleration and strain-rate system. **Fig. 1** illustrates the two-dimensional staggered grid configurations, detailing the spatial distribution of physical properties and wavefields for both formulations within an isotropic medium. These principles remain equally valid for rotated staggered grid geometries as well. Accordingly, the finite-difference discretization schemes originally developed for velocity and stress (e.g., Virieux 1986; Saenger and Bohlen 2004) can be readily adapted. For example, the discretization of the fourth equation in Eq. (9) is expressed as follows

$$[\dot{\varepsilon}_{xx}]_{i,j}^{n+1} = [\dot{\varepsilon}_{xx}]_{i,j}^{n} + \frac{\Delta t}{\Delta x}\left([a_x]_{i,j+\frac{1}{2}}^{n+\frac{1}{2}} - [a_x]_{i,j-\frac{1}{2}}^{n+\frac{1}{2}}\right), \tag{13}$$

where both temporal and spatial derivatives are discretized with second-order accuracy utilizing the symmetric leapfrog technique (Anderson and Wendt 1995). Here, $\Delta t$ and $\Delta x$ represent the discretization intervals in time and space, respectively. The index $n$ denotes the discrete time step, and $(i, j)$ indicate the spatial grid coordinates. The fractional indices, $1/2$, in the subscripts and superscripts represent the staggered half-steps inherent to the leapfrog scheme. Accordingly, the first equation in Eq. (12) is discretized as follows

$$[a_x]_{i,j+\frac{1}{2}}^{n+\frac{1}{2}} = \frac{1}{\rho_{i,j+\frac{1}{2}}}\frac{\Delta t}{\Delta x}\left([E]_{i,j+1}^{n} - [E]_{i,j-1}^{n}\right) + \frac{1}{\rho_{i,j+\frac{1}{2}}}\frac{\Delta t}{\Delta x}[F]_{i,j+\frac{1}{2}}^{n+\frac{1}{2}}, \tag{14}$$

where $E = c_{11}^{p}\dot{\varepsilon}_{xx} + c_{11}^{p}\dot{\varepsilon}_{yy} + c_{13}^{p}\dot{\varepsilon}_{zz}$ and $F = \partial f_x/\partial t$ . Furthermore, the treatment of physical properties remains consistent with the averaging techniques employed in the velocity-stress formulation, as detailed by Moczo et al. (2007). For example, the density at the half-step position in the $x$-direction is determined via arithmetic averaging, defined as

$$\rho_{i,j+\frac{1}{2}} = \frac{1}{2}\left(\rho_{i,j+1} + \rho_{i,j}\right). \tag{15}$$

In addition to discretization, other components of the finite-difference solution of the velocity and stress formulation including absorbing boundary conditions, dispersion properties, and numerical stability have been rigorously examined in the literature (e.g., Levander 1988; Graves 1996; Collino and Tsogka 2001; Ren and Liu 2015). Similar to discretization, these fundamental principles extend naturally to both acceleration and strain-rate formulation. Consequently, adapting existing seismic simulation tools to utilize acceleration and strain-rate rather than velocity and stress is technically seamless, requiring only minor algorithmic adjustments.

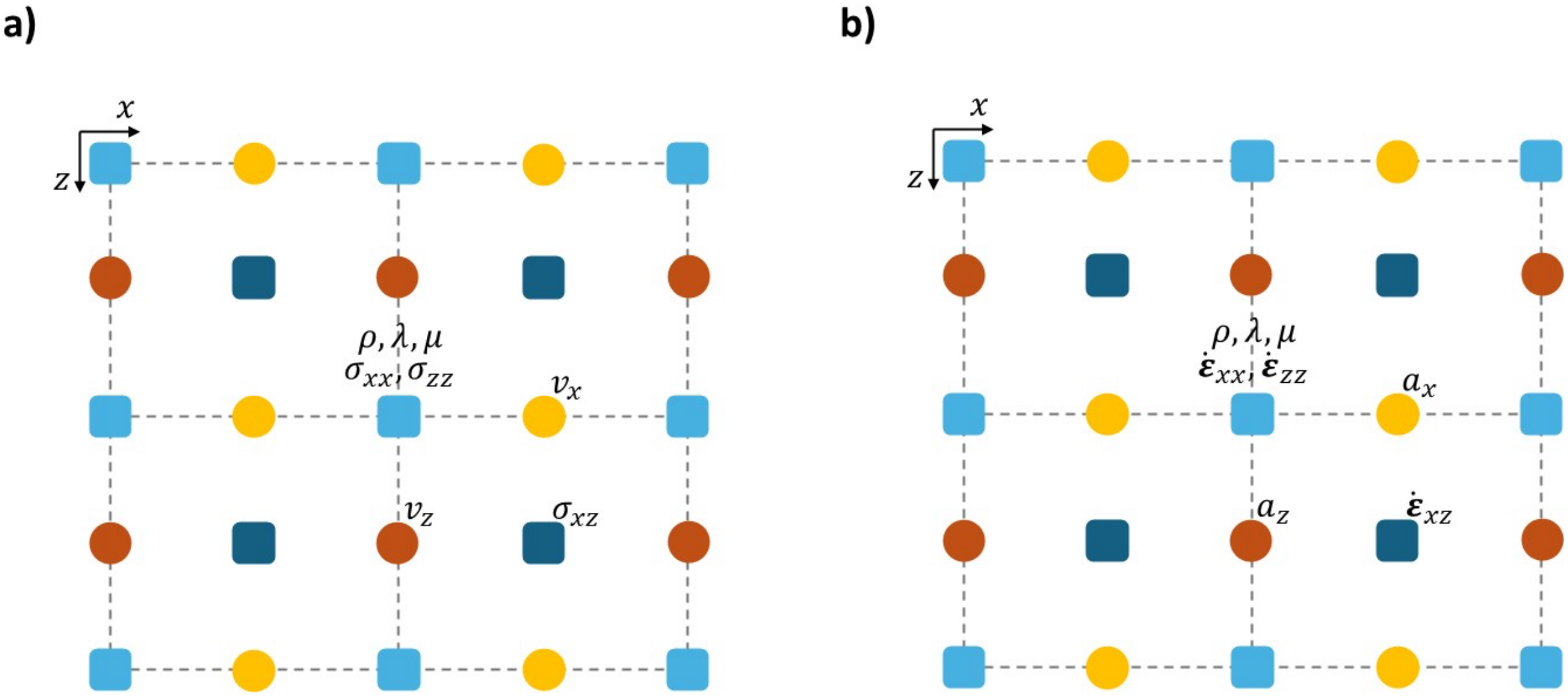

**Fig. 1** Staggered grids in two dimensions. The wavefields and physical properties distributed on the grids in a) Velocity and stress, and b) Acceleration and strain-rate formulations are very similar as a result of the similarity of the equations.

### 2.5. Seismic Response of Optical Fibre Cables

Fiber-optic cables, utilized in Distributed Acoustic Sensing (DAS), record the strain/strain-rate induced by seismic wavefields along the tangential axis of the cable at each sensing point. Consequently, as illustrated in **Fig. 2**, the strain-rate tensor computed within a global Cartesian coordinate system $(xyz)$ must be transformed into a local coordinate system $(x'y'z')$ aligned with the cable's orientation (Eaid et al. 2018). This spatial transformation is achieved by applying a rotation matrix to the strain-rate tensor as follows (Li et al. 2022)

$$\dot{\boldsymbol{\varepsilon}}_{x'y'z'} = \boldsymbol{R}\dot{\boldsymbol{\varepsilon}}_{xyz}\boldsymbol{R}^T, \quad (16)$$

where $\dot{\boldsymbol{\varepsilon}}_{x'y'z'}$ is the transformed strain-rate tensor, $\dot{\boldsymbol{\varepsilon}}_{xyz}$ is the original tensor, and $\boldsymbol{R}$ is the transformation matrix as following

$$\boldsymbol{R} = \begin{bmatrix} \boldsymbol{r}_{x'}.\boldsymbol{r}_x & \boldsymbol{r}_{x'}.\boldsymbol{r}_y & \boldsymbol{r}_{x'}.\boldsymbol{r}_z \\ \boldsymbol{r}_{y'}.\boldsymbol{r}_x & \boldsymbol{r}_{y'}.\boldsymbol{r}_y & \boldsymbol{r}_{y'}.\boldsymbol{r}_z \\ \boldsymbol{r}_{z'}.\boldsymbol{r}_x & \boldsymbol{r}_{z'}.\boldsymbol{r}_y & \boldsymbol{r}_{z'}.\boldsymbol{r}_z \end{bmatrix} \quad (17)$$

where $\boldsymbol{r}_i$ is the unit vector in the direction of $i$ axis.

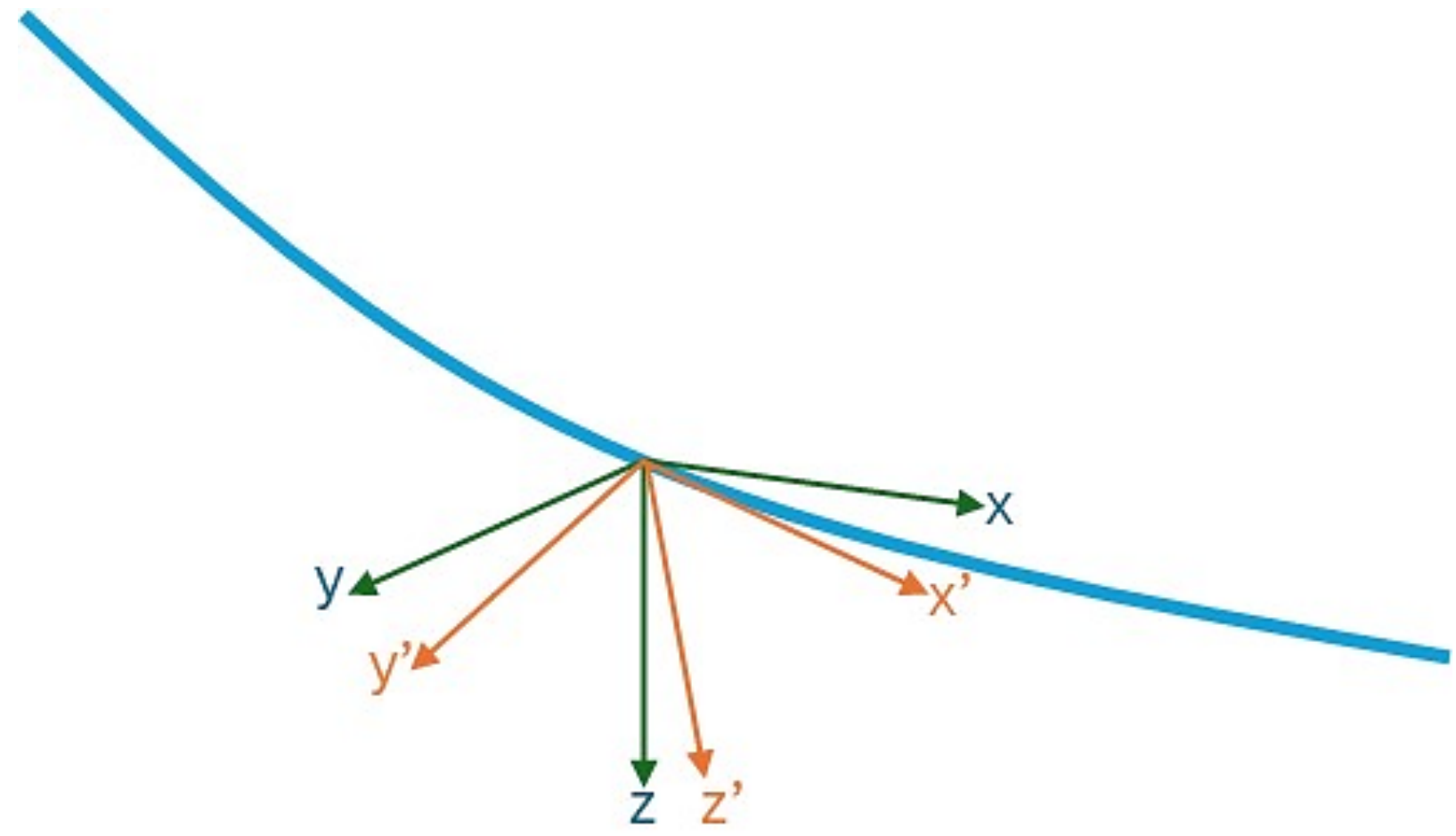

**Fig. 2** The global Cartesian coordinate (green axes) transformed to a local coordinate system (amber axes) for calculation of the tangential component.

## 3. Numerical Examples

This section presents an evaluation of the proposed elastic wave equation formulated in terms of acceleration and strain-rate. To assess the methodology, two numerical experiments are conducted in a two-dimensional domain. While 2D modelling is adopted for computational efficiency, the fundamental insights remain applicable to more general cases. The governing equations are discretized using a rotated staggered-grid finite-difference scheme (Saenger and Bohlen, 2004). This approach utilizes fourth-order spatial accuracy and second-order temporal accuracy. Both simulations employ a uniform square grid with a constant size of 5 m × 5 m. To mitigate boundary reflections and simulate an infinite medium, absorbing boundary conditions are implemented on all four sides of the model using a damping layer approach (Cerjan et al. 1985). The numerical analysis is structured to achieve the primary goal of benchmarking the accuracy of the acceleration and strain-rate formulation against the conventional velocity-stress equation. Additionally, the robustness and efficacy of the proposed formulation when characterizing wave propagation within complex media is demonstrated.

### 3.1. Layered Model

A two-layer model, measuring 2 km by 2 km, is utilized to benchmark the performance of the two formulations (**Fig. 3**). The source function is a Ricker wavelet with a 20 Hz peak frequency acting as a point source in the vertical direction. As illustrated in **Fig. 3**a, the source position is indicated by a red star, while the vertical and horizontal receiver arrays are represented by green and yellow stars, respectively. Numerical simulations were conducted using both formulations to facilitate a comparative analysis. **Fig. 4** presents a comparison of the strain-rate field snapshots at t = 0.55 s. The wavefields generated by both formulations exhibit excellent agreement, with a negligible relative difference on the order of $10^{-7}$, likely attributable to the minute difference in round-off errors of the respective finite-difference schemes. The four distinct wavefronts in the second layer are identified in Figures 4a-b. Given the source characteristics and the impedance contrast between layers, the PP and SS phases are the most prominent, while the converted PS and SP waves remain relatively weak. Notably, the shear strain-rate component is of a comparable magnitude to the normal components, indicating its significant influence on the overall response of the fiber-optic cables.

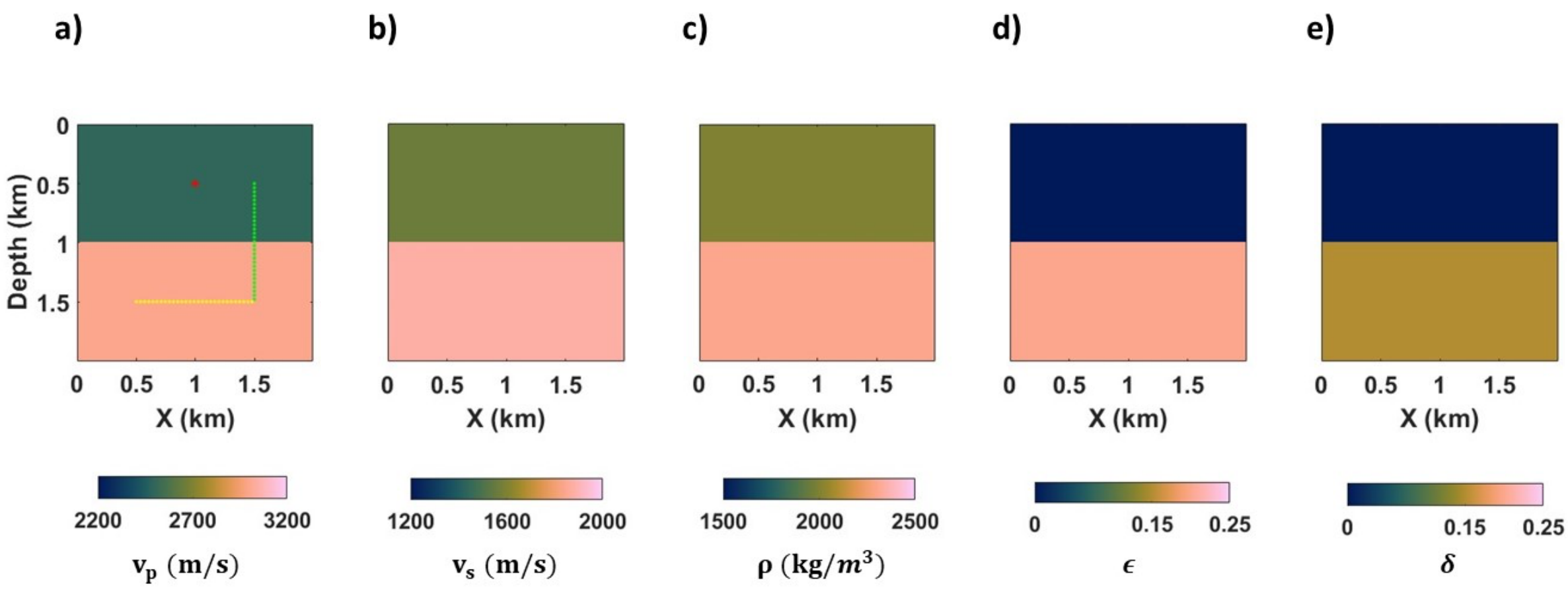

**Fig. 3** The a) P- and b) S-wave velocities, c) Density, and d) Epsilon and e) Delta parameters for the layered model. The red star denotes the source location, green stars show the vertical receiver line, and yellow stars denote the horizontal receiver spacing.

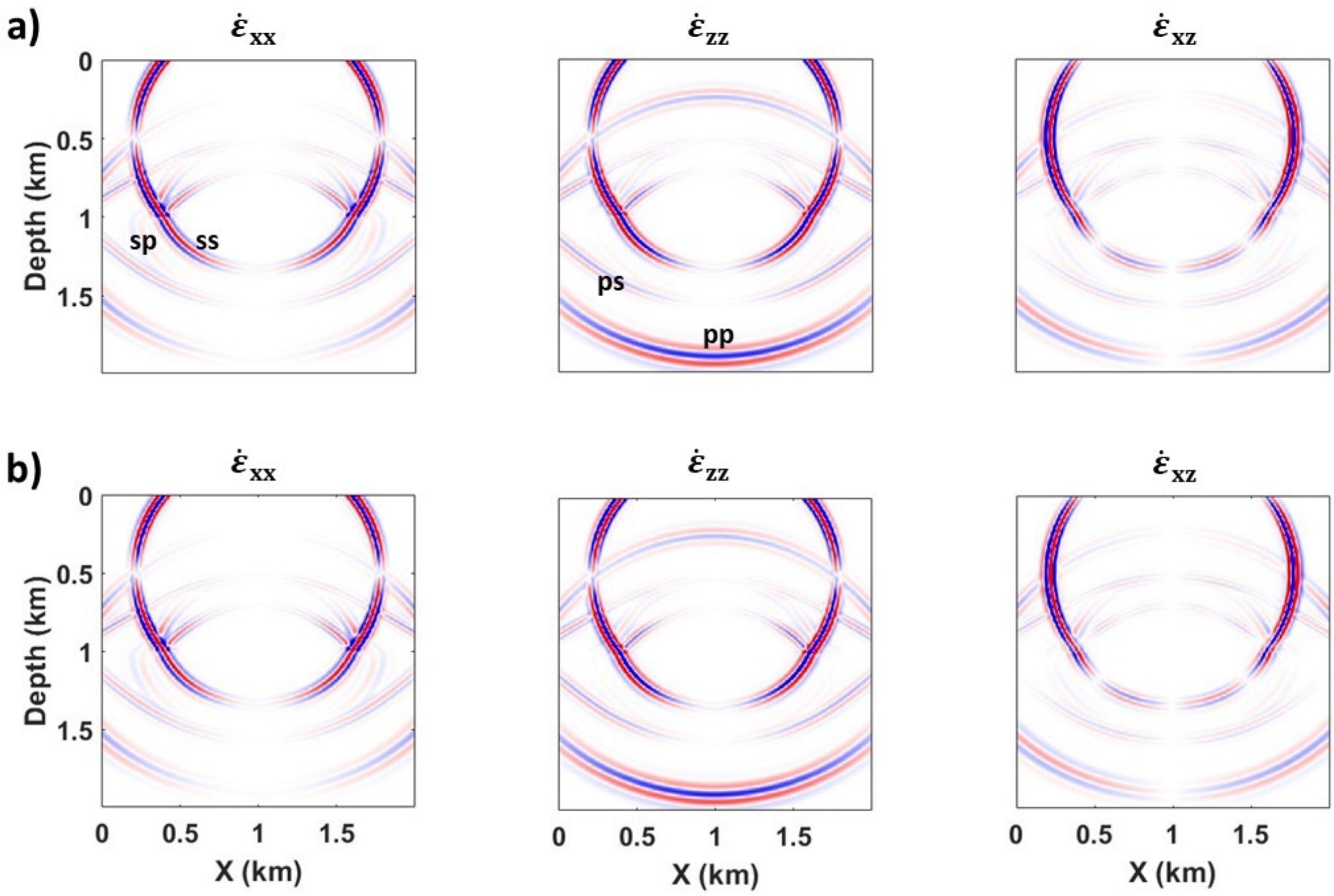

**Fig. 4** Snapshots of the strain-rate wavefields at 0.55 s for the layered model generated by a) Velocity and stress, and b) Acceleration and strain-rate formulations. The two formulations generate the same wavefields.

**Fig. 5** displays the recorded wavefields along the horizontal (**Fig. 5**a–b) and vertical (**Fig. 5**c–d) receiver arrays for both formulations. The horizontal axes, $X_r$ and $Z_r$, represent the receiver coordinates relative to the origin located at the top-left corner. To facilitate a more rigorous quantitative comparison, individual traces are extracted from each array at $X_r$ = 0.75 km and $Z_r$ = 0.75 km, as illustrated in **Fig. 6**. Both figures confirm that the two formulations generate equivalent wavefields with a high degree of precision, maintaining a negligible relative error of $10^{-7}$. As a result of the vertical orientation of the point source, P-wave energy is predominantly directed vertically, while S-waves propagate primarily in the lateral direction. In addition, converted PS and SP phases exhibit relatively low amplitudes. Consequently, the horizontal receiver line recordings are dominated by the first P-wave arrival, with other phases appearing significantly weaker. The wave types identified in **Fig. 4**a are similarly annotated in the recordings of **Fig. 5**a where the vertical receiver line records higher-amplitude S-wave arrivals, whereas the P-wave arrivals are comparatively exhibiting lower amplitude.

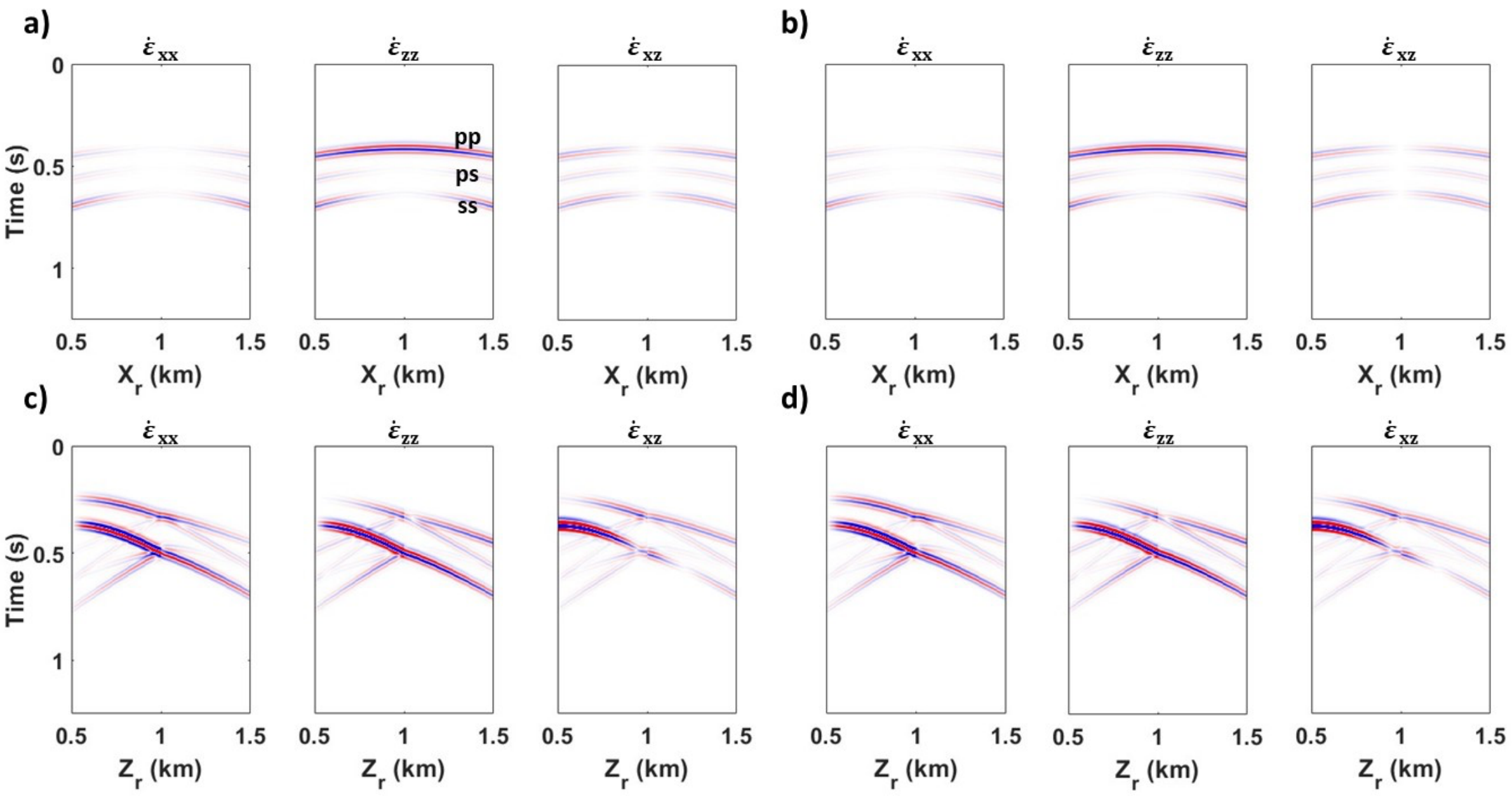

**Fig. 5** Recorded strain-rate components on the horizontal line using a) Velocity and stress, and b) Acceleration and strain-rate formulations. The same components on the vertical receiver line using c) Velocity and stress, and d) Acceleration and strain-rate formulations. The two formulations produce the same wavefields.

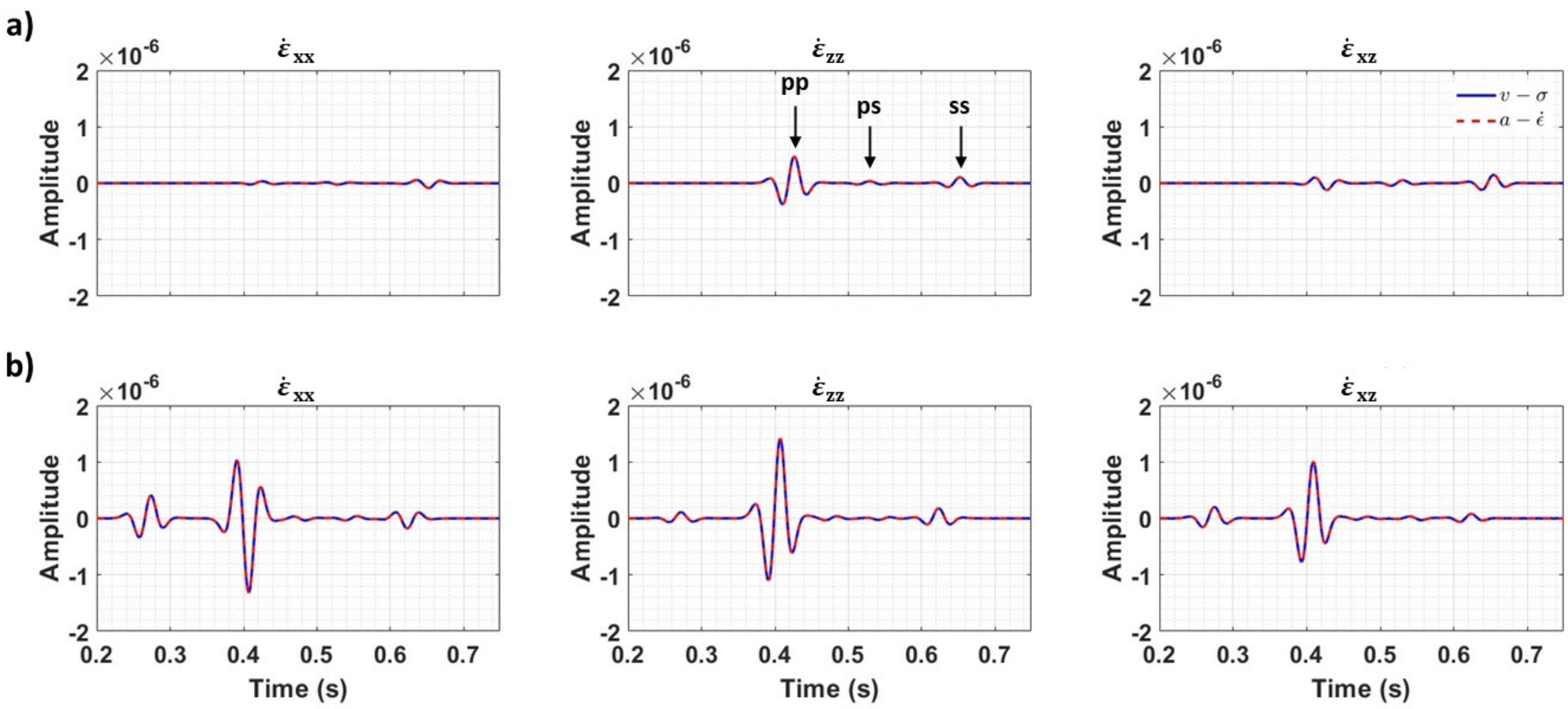

**Fig. 6** Recorded strain-rate components at $X_r$ = 0.75 km and $Z_r$ = 0.75 km on the a) Horizontal and b) Vertical recording lines. The two formulations generated the same wavefield with extremely negligible relative difference.

Following the elastic simulation, two acoustic simulations were conducted using the formulations in Eqs. (9) and (12). **Fig. 7** compares the resulting wavefields, demonstrating that the reflected and transmitted P-waves are identical across the normal components for both cases. According to Eq. (9), the shear strain-rate terms do not contribute to the acceleration vector update. Consequently, the normal strain-rates remain unaffected by shear components, leading Eqs. (9) and (12) to yield identical normal strain-rate results. The primary distinction, however, lies in the fact that Eq. (9) permits the generation of shear strains.

Although Eq. (9) does not eliminate the shear component, it unexpectedly produces stable shear wavefields despite its physical inconsistency. Such phenomena, where material deformation occurs without corresponding internal resistive forces, are characteristic of processes like thermal expansion or plastic deformation (Garmong, 1974), both of which diverge from the fundamental elastic assumptions underlying these equations. Hence, as discussed previously and demonstrated by this numerical experiment, the shear components must be explicitly nullified within the acceleration and strain-rate formulation to ensure the generation of physically consistent acoustic wavefields.

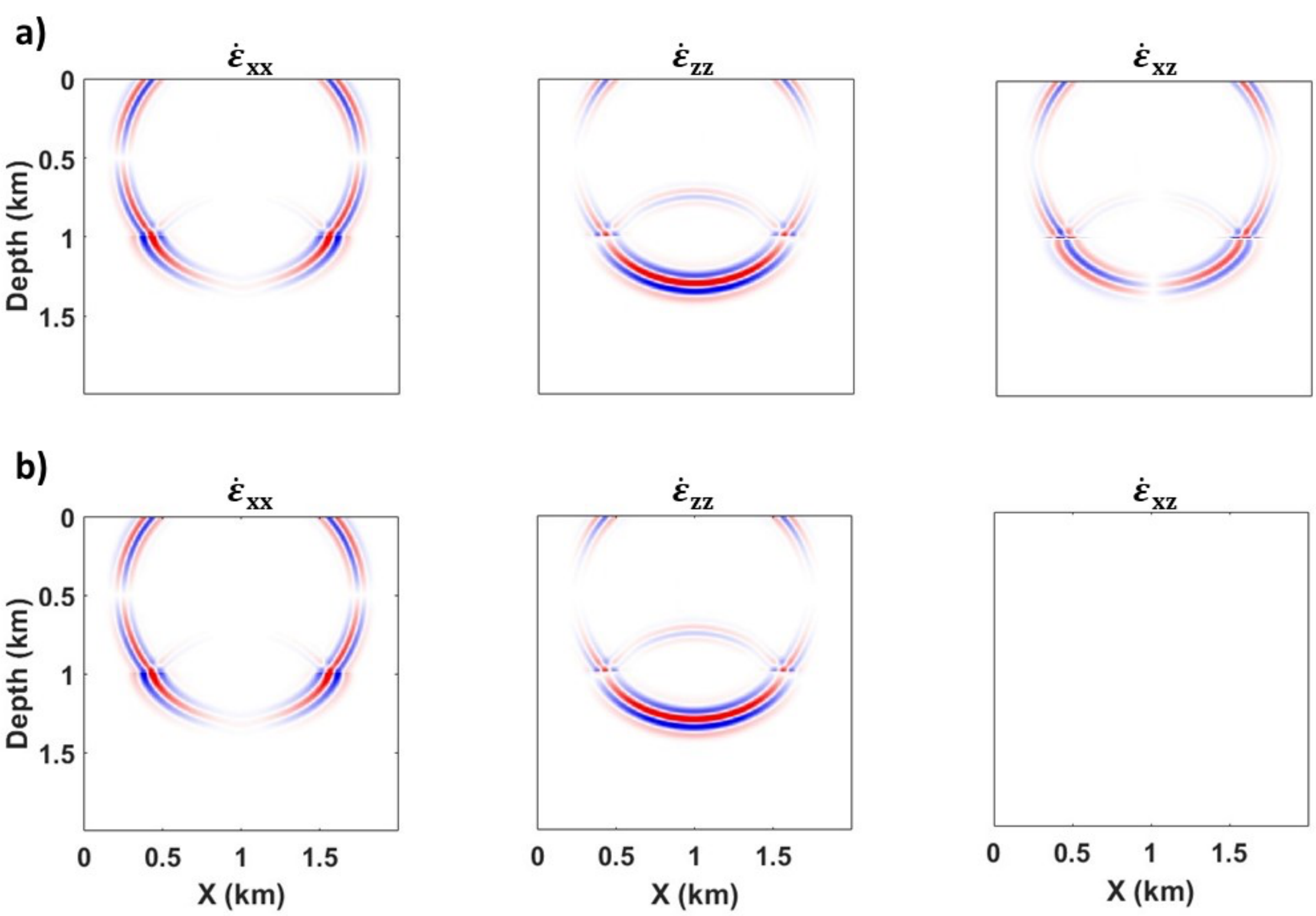

**Fig. 7** Acoustic simulations conducted using the formulations in a) Eq. (9), and b) Eq. (12). Equation (9) produces stable shear wavefields despite its physical inconsistency.

### 3.2. Marmousi Elastic Model

To evaluate the performance of the proposed formulation in complex media, numerical simulations are conducted using the elastic Marmousi model. For this experiment, the original Marmousi model is resampled and cropped to the dimensions illustrated in **Fig. 8**. A vertically oriented Ricker wavelet with a peak frequency of 10 Hz serves as the source function. The source position is indicated by a red star, and the receiver array is denoted by green stars (**Fig. 8**a). **Fig. 9** presents a comparison of the strain-rate field snapshots at t = 1.8 s, demonstrating that both formulations yield equivalent wavefields. Consistent with the previous layered model results, the relative difference between the two formulations remains negligible, on the order of $10^{-6}$. Furthermore, the shear strain-rate component is of a comparable magnitude to the normal components, highlighting its significant influence on the simulated response of fibre optic cables. Similar to the previous example, the vertical point source generates wavefields where P-wave energy is concentrated vertically, while S-wave energy propagates primarily in the lateral direction. As observed in the receiver array recordings, the primary P-wave arrivals on the vertical component exhibit maximum amplitude directly beneath the source, diminishing as the lateral offset increases (**Fig. 10**). Conversely, on the horizontal component, these same arrivals show minimal intensity directly below the source but gain more strength at larger offsets, consistent with the expected polarization of the wavefield. This experiment demonstrates that the acceleration and strain-rate formulation maintains numerical robustness in complex media, performing with the same efficacy and precision as the traditional velocity-stress formulation.

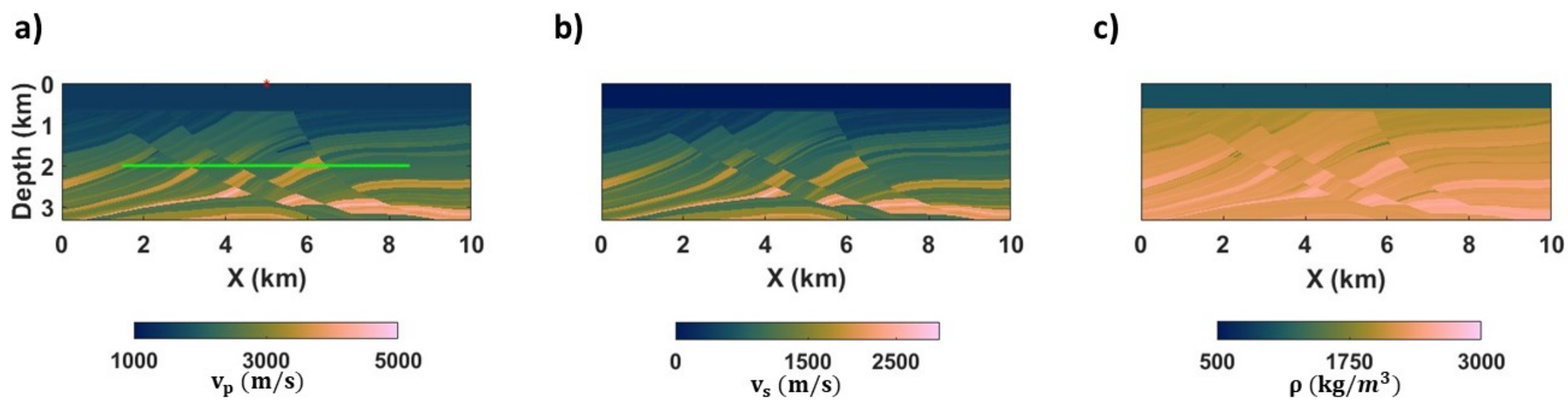

**Fig. 8** The a) P-wave and b) S-wave velocities, and c) Density from elastic Marmousi model. Red star shows the source position, and the receiver array is denoted by green stars.

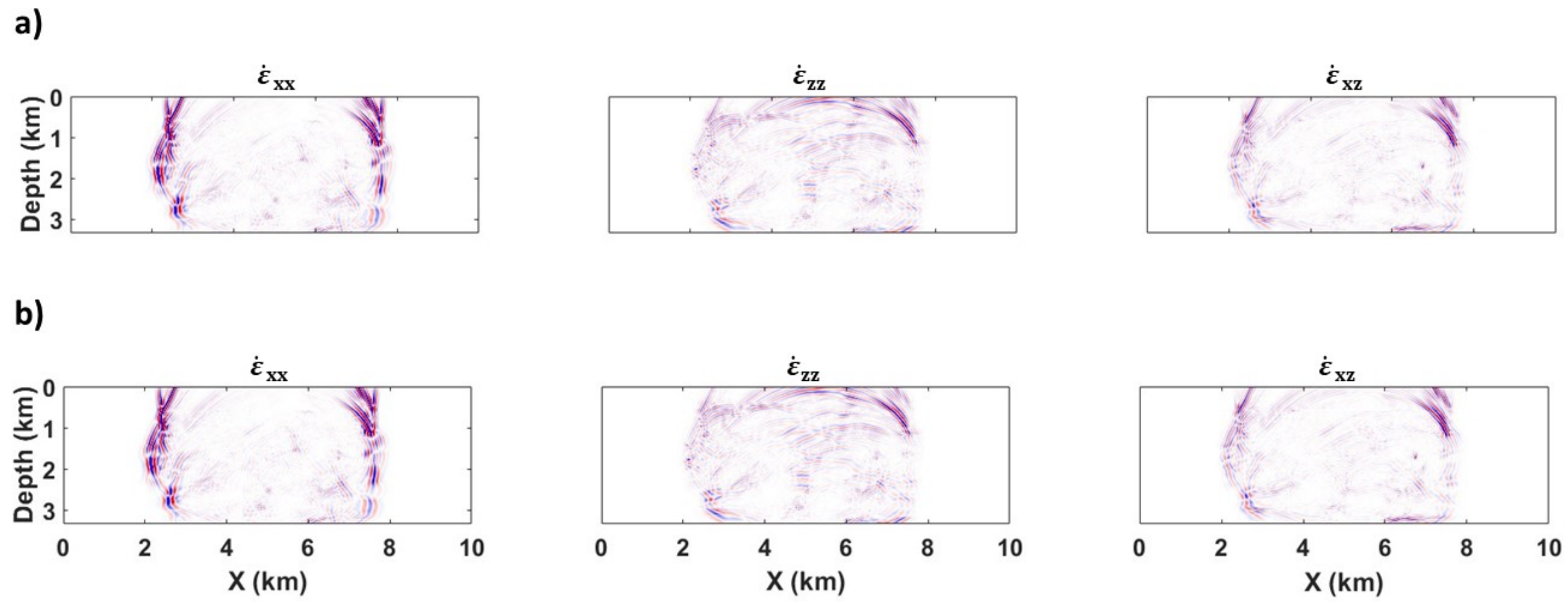

**Fig. 9** Comparing the wavefields generated by a) Velocity and stress, and b) Acceleration and stress formulations. The relative difference between the two formulations remains negligible, on the order of $10^{-6}$.

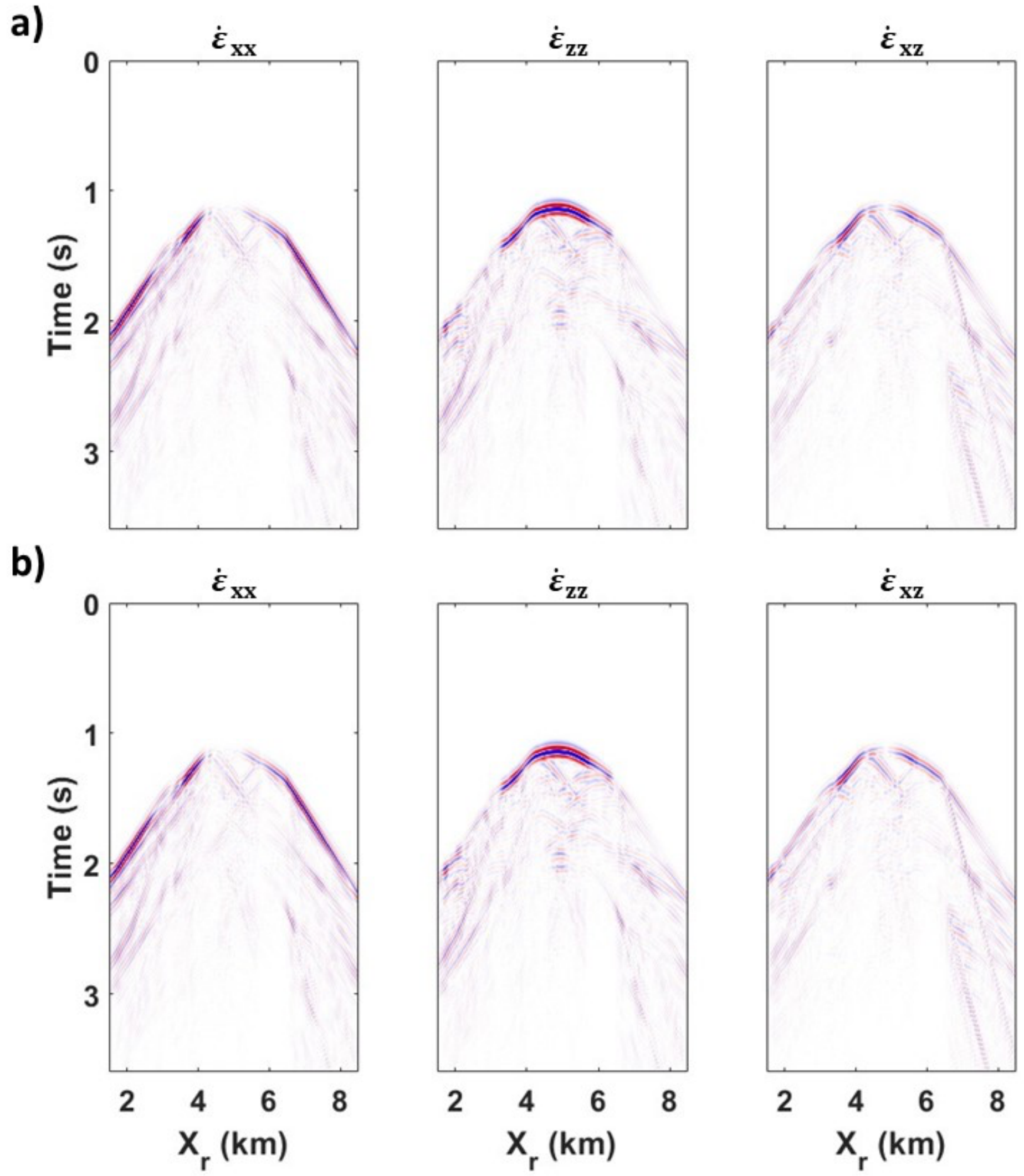

**Fig. 10** The receiver array recordings of a) Velocity and stress, and b) Acceleration and strain-rate wavefields. The recordings, consistent with the snapshots, confirm the same level of accuracy between the two formulations.

## 4. Conclusions

The proposed acceleration and strain-rate formulation for elastic wave propagation is systematically derived within a framework analogous to the conventional velocity-stress formulation. Consequently, it preserves a consistent mathematical structure, comprising nine first-order partial differential equations. This elastic framework is further simplified for acoustic media, offering a computationally efficient alternative for specific seismic applications. Finite-difference discretization demonstrates that the proposed system is compatible with the gridding and numerical techniques standard to velocity-stress implementations. Extensive numerical experiments confirm that the acceleration and strain-rate formulation achieves a level of accuracy equivalent to the velocity and stress formulation, even in the case of highly complex media. Given its seamless integration and ability to eliminate the need for data conversion or extra computations, this formulation provides a robust and efficient alternative for Distributed Acoustic Sensing (DAS) seismic simulation and imaging.

## 5. Appendix A: Stiffness Tensor in VTI Media

The stiffness tensor for vertically transverse isotropic medium is as following

$$C = \begin{bmatrix} c_{11} & c_{12} & c_{13} & 0 & 0 & 0 \\ c_{12} & c_{11} & c_{13} & 0 & 0 & 0 \\ c_{13} & c_{13} & c_{33} & 0 & 0 & 0 \\ 0 & 0 & 0 & c_{44} & 0 & 0 \\ 0 & 0 & 0 & 0 & c_{44} & 0 \\ 0 & 0 & 0 & 0 & 0 & c_{66} \end{bmatrix}, \tag{A1}$$

where

$$c_{12} = c_{11} - 2c_{66}. \tag{A2}$$

The components of the tensor, in terms of Thomsen's parameters are as follows (Thomsen 1986; Tsvankin 1996)

$$\begin{aligned} c_{11} &= \rho(1+2\epsilon)v_p^2, \\ c_{12} &= \rho(1+2\epsilon)v_p^2 - 2\rho(1+2\gamma)v_s^2, \\ c_{13} &= \rho\sqrt{f(f+2\delta)}v_p^2 - \rho v_s^2, \\ c_{33} &= \rho v_p^2, \\ c_{44} &= \rho v_s^2, \\ c_{66} &= \rho(1+2\gamma)v_s^2, \end{aligned} \tag{A3}$$

where

$$f = 1 - \left(\frac{v_s}{v_p}\right)^2. \tag{A4}$$

The components of the tensor in the acoustic media are as following

$$\begin{aligned} c_{11}^p &= c_{12}^p = \rho(1+2\epsilon)v_p^2, \\ c_{13}^p &= \rho\sqrt{1+2\delta}v_p^2, \\ c_{33}^p &= \rho v_p^2, \\ c_{44}^p &= c_{66}^p = 0. \end{aligned} \tag{A5}$$